# A relation between the multiplicity of the second eigenvalue of a graph Laplacian, Courant's nodal line theorem and the substantial dimension of tight polyhedral surfaces

Tsvi Tlusty

*Department of Physics of Complex Systems, Weizmann Institute of Science, Rehovot 76100, Israel*

**Abstract.** This note discusses a relation between the multiplicity $m$ of the second eigenvalue $\lambda_2$ of a Laplacian on a graph $G$, tight mappings of $G$ and a discrete analogue of Courant's nodal line theorem. For a certain class of graphs, we show that the $m$-dimensional eigenspace of $\lambda_2$ is tight and thus defines a tight mapping of $G$ into an $m$-dimensional Euclidean space. The tightness of the mapping is shown to set Colin de Verdière's upper bound on the maximal $\lambda_2$-multiplicity, $m \leq chr(\gamma(G)) - 1$, where $chr(\gamma(G))$ is the chromatic number and $\gamma(G)$ is the genus of $G$.

## 1. Introduction

Laplacians, discrete or continuous, are omnipresent in physics and mathematics and much effort has been directed to analysis of their spectra [1-5]. A seminal result in this field is Courant's nodal line theorem that relates the order of the Laplacian's eigenvalues to the sign patterns of the corresponding eigenfunctions [6, 7]: *If the eigenfunctions of a Laplacian on a domain are ordered according to increasing eigenvalues, then the nodes of the n-th eigenfunction divide the domain into no more than n nodal domains.* The *nodal domains*, in which the eigenfunction takes one sign, are separated by the nodal sets that are the zero level-sets of the eigenfunction. For graph Laplacians, the discrete analogue of the nodal domain becomes the *sign-graph*, a maximal connected subgraph on which an eigenfunction takes the same sign. On *weak* sign-graphs the eigenfunction is either non-positive or non-negative, while on *strong* sign-graphs the sign of the eigenfunction is strictly positive or negative. Several authors have found the following discrete analogue of Courant's nodal line theorem [5, 8-10]:

**Theorem 1.** *On a connected graph G, the n-th eigenfunction* $\mathbf{u}_n$ *of the Laplacian* $\Delta$ *has at most n weak sign-graphs.*

For the special case of the second eigenvalue $\lambda_2$, it was shown earlier that the corresponding eigenfunction $\mathbf{u_2}$ cuts the graph into exactly two weak sign-graphs [11, 12]. Among the eigenvalues of the graph Laplacian $\lambda_2$ plays a special role. Physically, it corresponds to the mixing time of a Brownian walk on the graph and to its first-excited energy level. Mathematically, $\lambda_2$ was shown to be related to several structural and isoperimetric properties of the graph, such as the max-cut problem [13-15].

Of special interest is $m$, the multiplicity of $\lambda_2$ and the dimension of the $\lambda_2$-eigenspace. $m$ is the degeneracy of the first-excited energy level, which is the first mode to appear at several types of continuous (or second-order) phase transitions. Continuous phase transitions are known to occur in noisy information channels when the signal distortion level is varied [16] and have been related to certain classification and optimization problems [17]. We suggested that this type of phase transition may occur during the Darwinian evolution of noisy biological information channels, and may be the mechanism underlying the emergence of codes in these channels [18]. Noisy information channels may be described in terms of error-graphs, in which edges connect two signals that are likely to be confused by noise. The Laplacian of the error-graph is the operator that measures the average effect of errors and therefore controls the phase transition. This motivated the present note that focuses on the relation between the topology of the error-graph and the maximal degeneracy of the Laplacian's first-excited modes.

Colin de Verdière revealed an intimate relation between the topology of a Riemannian surface $S$ and the supremum of $m$ over all possible Laplacians on the surface $\overline{m}(S)$ through the chromatic number [4, 5, 19, 20]:

**Theorem 2.** *For any surface S, the maximal multiplicity of the Laplacian's second eigenvalue is bounded from below,* $\overline{m}(S) \geq chr(S) - 1$.

The chromatic number of a surface $chr(S)$ is defined as the least number of colors required to color any map on $S$. The dependence of the chromatic number on the topology of $S$ is given by Heawood's formula [21], $chr(S) = \left\lfloor \left(7 + \sqrt{49 - 24\chi(S)}\right)/2 \right\rfloor$, where $\chi(S)$ is the Euler characteristic of the surface (the genus of an oriented surface is $\gamma(S) = 1-\chi(S)/2$). Colin de Verdière conjectured that the lower bound of Theorem 2 is exactly the maximal possible multiplicity, $\overline{m}(S) = chr(S) - 1$ [20].

The multiplicity $\overline{m}(S)$ was shown to be bounded also from above [22]. In a series of improvements [23-25], this upper bound approached the lower bound, $\overline{m}(S) \leq 5 - \chi(S)$. The last improvement is due to Séveneec, who also brought up a potential link between the multiplicity $\overline{m}(S)$ of Laplacians on a surface $S$ and *tight* mappings of $S$ into Euclidean spaces. These mappings are termed tight because no hyperplane can cut the mapped surface into more than two pieces, which gave tightness its other name, the *two-piece property*. Banchoff showed that if $S$ is a polyhedral surface, then the substantial dimension of such tight mappings $d$ is bounded , $d \leq chr(S) - 1$ [26].

The equality of Banchoff's and Colin de Verdière's bounds is not a mere coincidence. The purpose of the present note is to elucidate this relation between tight mappings, the maximal multiplicity of the second eigenvalue of graph Laplacians and Courant's nodal line theorem, and to use this relation to prove that Colin de Verdière's bound applies for a rather wide class of graphs (Theorem 3).

In section 2 we list the basic notation and discuss the relation between tightness and the nodal pattern of the $\lambda_2$-eigenfunctions. We show that the $\lambda_2$-eigenspace of certain graphs, namely paths, cycles, complete graphs and their Cartesian products are tight. In section 3 we prove Colin de Verdière's bound for the class of graphs whose $\lambda_2$-eigenspaces are tight. The sketch of the simple proof is the following: We consider a graph $G$ that is embedded into a polyhedral surface $S$. The $\lambda_2$-eigenspace of the graph Laplacian is $m$-dimensional and defines a tight mapping of $G$ and $S$ into the $m$-dimensional Euclidean space. This sets Banchoff's upper bound on the dimension of the Euclidean space, which by the construction of the mapping is also the maximal multiplicity, $m \leq chr(S) - 1 = chr(\gamma(G)) - 1$.

## 2. Tightness and the $\lambda_2$-eigenfunctions of a Laplacian

**Basic notations:** Before we discuss the question of which $\lambda_2$-eigenspaces are tight, we list some basic notations related to the graph Laplacian and to the mapping induced by its eigenfunctions. We consider an undirected, simple, loop-free graph $G$ with a vertex set $V$ of $N$ vertices and an edge set $E$. The vertices are denoted by $i$ = 1, 2…N. The graph *Laplacian* $\Delta$ is an operator which may be expressed as a real symmetric matrix $\Delta \in \boldsymbol{R}^{N\times N}$ associated with the graph in the following way: If vertices $i$ and $j$ are adjacent, $(i, j) \in E$, then the corresponding entry in the Laplacian is negative, $\Delta_{ij} = \Delta_{ji} < 0$, otherwise $\Delta_{ij} = 0$, and the diagonal terms assure that the sum over rows and columns vanishes, $\Delta_{ii} = -\sum_{j\neq i}\Delta_{ij}$. The operator $\Delta$ is sometimes termed a *weighted Laplacian* while 'Laplacian' is reserved for the case where the negative entries are all $\Delta_{ij} = -1$. $\Delta$ is irreducible iff the associated graph is connected. By the Perron-Frobenius theorem, the first eigenvalue $\lambda_1 = 0$ is non-degenerate and the $\lambda_1$-eigenfunction may be chosen to be the all-ones vector **1**. The multiplicity of the second eigenvalue $\lambda_2$ is denoted by $m$. Similar to the continuous case, one may define the maximal $\lambda_2$-multiplicity over all possible Laplacians on a graph, which is denoted by $\bar{m}(G)$.

Every graph $G$ can be *embedded* in a surface $S$ by a continuous one-to-one mapping (homeomorphism) $i$: $G \to S$. Intuitively, this means that $G$ can be drawn on $S$ with no intersecting edges. The *genus of a graph* $\gamma(G)$ is the minimal genus of a surface $S$ into which $G$ can be embedded ([27], chapter 3). The surface $S$ can in turn be embedded into a Euclidean space $\boldsymbol{R}^d$ by another homeomorphism $\varphi$: $S \to \boldsymbol{R}^d$. A *polyhedral surface* is a surface embedded in $\boldsymbol{R}^d$ such that its image $\varphi(S)$ is a finite union of planar polygons, termed *faces.* If $\varphi$ is only locally one-to-one it defines a *polyhedral immersion*, in which the polygonal faces can intersect. An embedded or immersed surface $S$ is *substantial* if it is not contained in any hyperplane. An embedded surface $S$ is *tight* if its intersection with any half-space $S\cup h$ is connected. Similarly, an immersion $S$ is tight if the preimage of the intersection with any half-space $\varphi^{-1}(S\cup h)$ is connected ([28], chapter 2). Functions (or vectors) on a graph are also maps, from the vertices into the real numbers, **u**: $V \rightarrow \boldsymbol{R}$ and we can therefore apply a similar notion of tightness:

**Definition 1.** *Let* **u** *be a function on a graph G and* $s \in \boldsymbol{R}$ *a level. Let* $G_+(\mathbf{u}, s)$ *and* $G_-(\mathbf{u}, s)$ *the subgraphs induced by the vertex sets* $V_+(\mathbf{u}, s) = \{i \in V \mid \mathbf{u}(i) \geq s\}$ *and* $V_-(\mathbf{u}, s) = \{ i \in V \mid \mathbf{u}(i) \leq s\}$*, respectively. If for any level* $s \in \boldsymbol{R}$ *both* $G_+(\mathbf{u}, s)$ *and* $G_-(\mathbf{u}, s)$ *are connected (or empty), then the function* **u** *is tight. A vector space of functions is tight if it contains only tight functions.*

The intuitive link between tightness and $\lambda_2$-eigenfunctions is evident. $\lambda_2$-eigenfunctions cut graphs into two weak sign-graphs, one non-negative and one non-positive, while tight surfaces are cut by hyperplanes into no more than two pieces. Trying to apply this intuition, one notes an obstacle: Nodal domains are determined by zero level-sets, that is by hyperplanes that pass through the origin 0, while tightness is a somewhat stricter condition on every affine hyperplane, not only those that pass through the origin. This difference is demonstrated in Figure 1, where two functions on the cycle graph $C_{20}$ are plotted. Both functions are cut into two sign-graphs by the zero-level set, and can therefore be, in principle, $\lambda_2$-eigenfunctions of a Laplacian on $C_{20}$. However, one of the functions is not tight, because it can be cut by certain nonzero level-sets into more than two pieces. In this simple example, tightness was lost by the presence of more than two extrema.

An immediate corollary of Definition 1 is that any function **u** on a complete graph $K_n$ is tight, since all the vertices are connected to each other. It is evident from Definition 1 that a tight function can have only one minimum and one maximum (Figure 1). Using matrix reducibility arguments, Fiedler showed that $\lambda_2$-eigenfunctions of graph Laplacians have a property related to tightness but somewhat weaker ([12], Theorem (3,3)):

**Lemma 1**. *If* **u** *is a* $\lambda_2$*-eigenfunction of a Laplacian of a connected graph G then*

(a) $G_+(\mathbf{u}, s)$ *is connected for any* $s \leq 0$ (b) $G_-(\mathbf{u}, s)$ *is connected for any* $s \geq 0$.

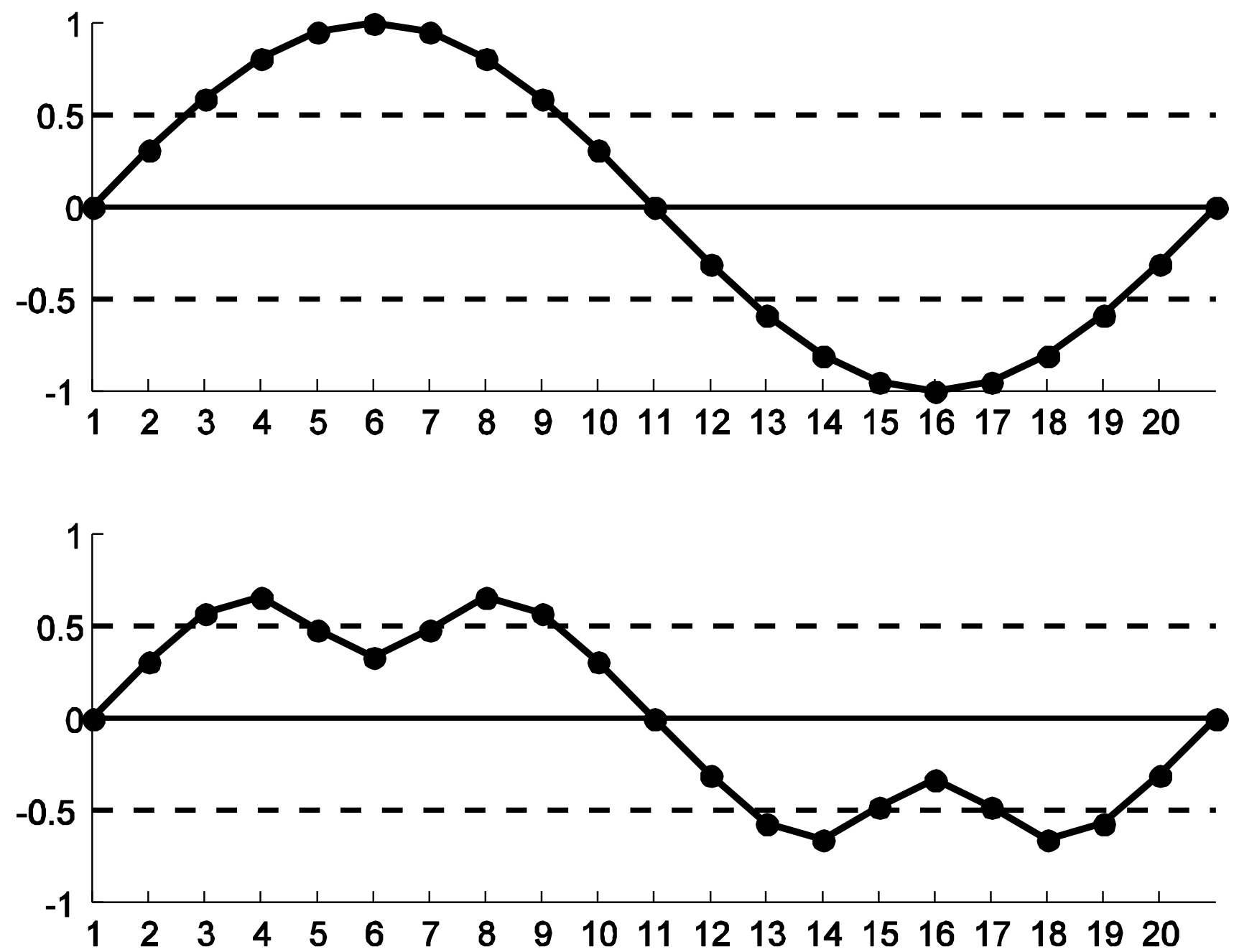

**Fig. 1.** Two functions on the cycle graph $C_{20}$. The $x$ coordinate specifies the vertex and the $y$ coordinate is the value of the function at this vertex. Both functions are cut into two sign-graphs by the zero-level set ($s = 0$, solid line) and could be $\lambda_2$-eigenfunctions by Theorem 1. However, the lower function is not tight, because it is cut by several level-sets (e.g. $s = \pm ½$, dashed lines) into more than two pieces.

We use Lemma 1 to prove that $\lambda_2$-eigenspaces of cycles and paths are always tight.

**Lemma 2**. *If the maximal degree of a connected graph G is 2, that is G is a cycle or a path, then the $\lambda_2$-eigenspace of a Laplacian on G is tight.*

**Proof**. Assume that a $\lambda_2$-eigenfunction $\mathbf{u}$ is not tight. This means that for some $s$ one of the induced subgraphs $G_+(\mathbf{u}, s)$ and $G_-(\mathbf{u}, s)$ is not connected. Without loss of generality, assume that $s > 0$ and by Lemma 1 $G_-(\mathbf{u}, s)$ is a connected path. If $G$ is a cycle then $G_+(\mathbf{u}, s)$ must also be connected and $\mathbf{u}$ is therefore tight (Figure 2, left).

If $G$ is a path, then $G_-(\mathbf{u}, s)$ separates the two components of $G_+(\mathbf{u}, s)$ ($G_+(\mathbf{u}, s)$ cannot have more than two components since $G_-(\mathbf{u}, s)$ has no more than two edges at its boundary). By the definition of the induced subgraphs, $G_+(\mathbf{u}, s) \subseteq G_+(\mathbf{u}, 0)$ and $G_-(\mathbf{u}, 0) \subseteq G_-(\mathbf{u}, s)$. By Theorem 1 $G_+(\mathbf{u}, 0)$ is connected and therefore must contain $G_-(\mathbf{u}, s)$ (Figure 2 right). It follows that $G_-(\mathbf{u}, 0) \subseteq G_-(\mathbf{u}, s) \subseteq G_+(\mathbf{u}, 0)$. $G_-(\mathbf{u}, 0)$ and $V_-(\mathbf{u}, 0)$ must therefore be empty. However, since **u** is orthogonal to the $\lambda_1$-eigenfunction, **1**, it must contain negative entries and $G_-(\mathbf{u}, 0)$ is nonempty. The eigenfunction **u** must therefore be tight. □

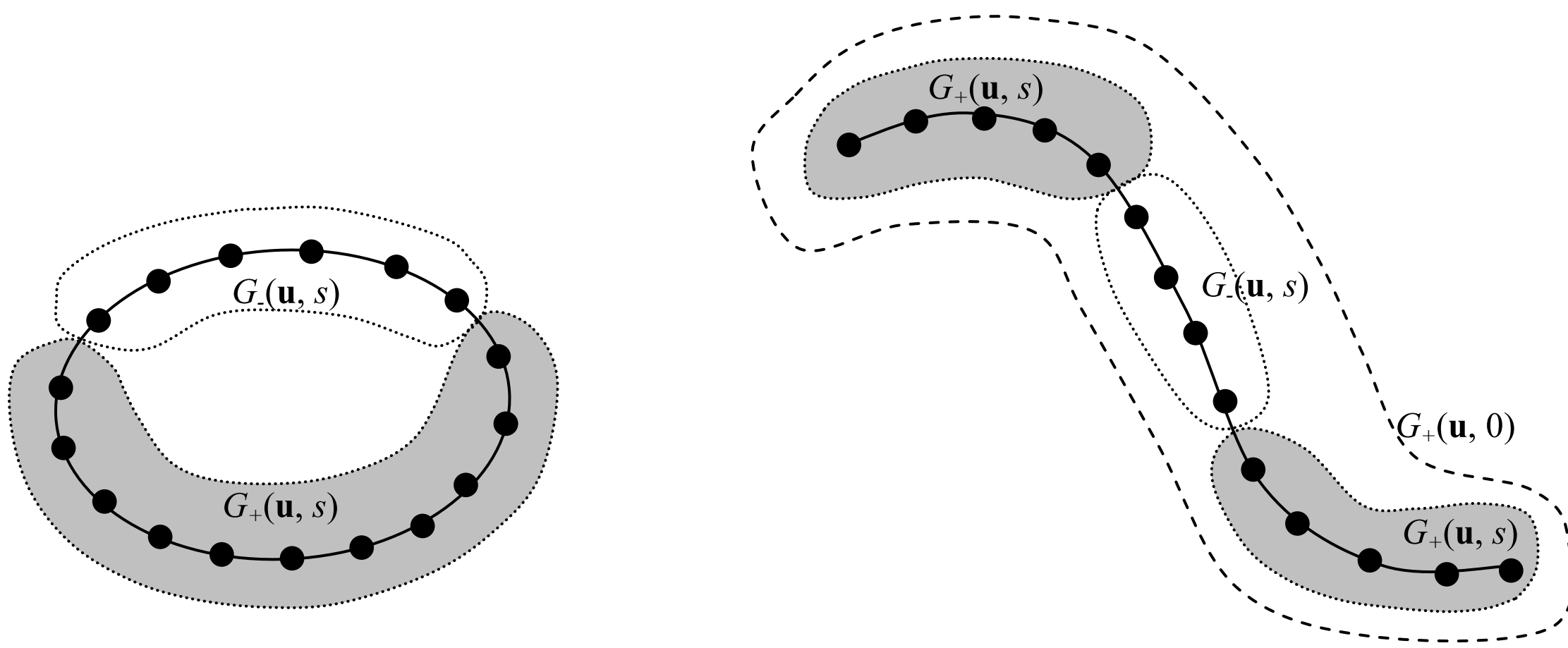

Fig. 2. The partition of a cycle (left) and a path (right) by the subgraphs induced by a level $s > 0$ and the $\lambda_2$-eigenfunction **u** (proof of Lemma 2). The induced subgraph $G_-(\mathbf{u}, s)$ is connected by Lemma 1. If $G$ is a cycle then $G_+(\mathbf{u}, s)$ must also be connected and **u** is therefore tight (left). If $G$ is a path, then by Courant's Theorem 1 $G_+(\mathbf{u}, 0)$ is connected and therefore must contain $G_-(\mathbf{u}, s)$ (right). It follows that the eigenfunction **u** is tight (see proof).

Returning to our previous example in Figure 1, it follows from Lemma 2 that the lower function that is not tight cannot be a $\lambda_2$-eigenfunction of a Laplacian on the cycle graph, although it has only two sign-graphs. The presence of a minimum between the two maxima allows the $s = ½$ level set to cut $G_-(\mathbf{u}, ½)$ into two components, thus contradicting Lemma 1. Similarly, the $s = -½$ level-set cuts the $G_+(\mathbf{u}, ½)$ into two components, again contradicting Lemma 1.

The only possible critical points on paths and cycles are maxima and minima. This is because these graphs are inherently unidirectional and to identify a saddle point at a vertex one needs at least two directions and more than two neighbors. A saddle point could separate the two components of $G_+(\mathbf{u}, s > 0)$ keeping $G_-(\mathbf{u}, s > 0)$ connected (but not simply-connected). Thus, the presence of a saddle point allows a $\lambda_2$-eigenfunction that is not tight. We remark that if the graph is a 1-skeleton of a polyhedral surface then the numbers of critical points are related by Morse's formula [29], $|\text{maxima}| - |\text{saddles}| + |\text{minima}| = \chi = 2 - 2\gamma$ . Therefore, for tight functions $|\text{maxima}| = |\text{minima}| = 1$ and $|\text{saddles}| = 2\gamma$.

Cartesian graph products appear in various fields, such as coding and information theory. The *Cartesian product* $P = G\Box H$ of the two graphs $G$ and $H$ has the vertex set $V_P = V_G \otimes V_P = \{(g, h) \mid g\in V_G, h\in V_H\}$, that is all the ordered pairs of vertices from $G$ and $H$. Two vertices $(g_1, h_1)$ $(g_2, h_2)$ are adjacent iff $g_1 = g_2$ and $h_1$, $h_2$ are adjacent in $H$, or $h_1 = h_2$ and $g_1$, $g_2$ are adjacent in $G$. The Laplacian of the graph product is given by $\Delta_P = \Delta_G \otimes I_H + I_G \otimes \Delta_H$, where $\otimes$ is the Kronecker tensor product and $I_G$, $I_H$ are the identity matrices of $G$ and $H$, respectively. It follows that the eigenfunctions of $\Delta_P$ are also products of the eigenfunctions of $\Delta_G$ and $\Delta_H$, $\mathbf{u}_P = \mathbf{u}_G \otimes \mathbf{u}_H$ with the corresponding eigenvalues $\lambda_P = \lambda_G + \lambda_H$ [30]. The $\lambda_2$-eigenfunctions of the Cartesian graph product $\mathbf{u}_P$ are either $\mathbf{u}_P = \mathbf{1}_G \otimes \mathbf{u}_H$ with the second eigenvalue $\lambda_{2,P} = \lambda_{2,H}$ or $\mathbf{u}_P = \mathbf{u}_G \otimes \mathbf{1}_H$ with $\lambda_{2,P} = \lambda_{2,G}$, where $\mathbf{1}_G$ and $\mathbf{1}_H$ are the all-ones vectors on $G$ and $H$, respectively. In case of degeneracy $\lambda_{2,P} = \lambda_{2,H} = \lambda_{2,G}$ and both function products are $\lambda_2$-eigenfunctions of $P = G\Box H$.

**Lemma 3.** *Let $P = G\Box H$ be the Cartesian graph product of the graphs $G$ and $H$. If the $\lambda_2$-eigenspaces of $G$ and $H$ are tight, then the $\lambda_2$-eigenspace of $P$ is tight.*

**Proof.** Without loss of generality assume that $\lambda_{2,G} < \lambda_{2,H}$ so that the $\lambda_2$-eigenfunctions of $P = G\Box H$ are of the form $\mathbf{u}_P = \mathbf{u}_G \otimes \mathbf{1}_H$. The entry of $\mathbf{u}_P$ at the vertex $(g, h)$ is therefore $\mathbf{u}_P(g, h) = \mathbf{u}_G(g)$. Assume that $\mathbf{u}_P$ is not tight, then for some $s$ one of the induced subgraphs $P_+(\mathbf{u}_P, s)$ and $P_-(\mathbf{u}_P, s)$ is not connected (the other one is connected by Lemma 1). Without loss of generality assume that $s > 0$, so that the induced subgraph $P_+(\mathbf{u}_P, s)$ is not connected. We now examine $G_+(\mathbf{u}_G, s)$, the subgraph induced by the same level $s$ on $G$. From the construction of $P = G\Box H$ and its $\lambda_2$-eigenfunction $\mathbf{u}_P = \mathbf{u}_G \otimes \mathbf{1}_H$ follows that if a vertex of $P$ belongs to the induced subgraph, $(g, h) \in P_+(\mathbf{u}_P, s)$, then the corresponding vertex of $G$ belongs to the induced subgraph, $g \in G_+(\mathbf{u}_G, s)$. Similarly, if $(g, h) \in P_-(\mathbf{u}_P, s)$ then $g \in G_-(\mathbf{u}_G, s)$. An immediate consequence is that $G_+(\mathbf{u}_G, s)$ is also not connected since it can be connected only through vertices $g$ such that belong to $G_-(\mathbf{u}_G, s)$, contradicting our assumption that $\mathbf{u}_G$ is tight. $\square$

The intuition underlying the proof is illustrated in Figure 3, where a $\lambda_2$-eigenfunction $\mathbf{u}_P = \mathbf{1} \otimes \mathbf{u}_{C20}$ of the Cartesian product of the cycles $P = C_{10} \square C_{20}$ is plotted. The projection of the eigenfunction on $C_{20}$ is $\mathbf{u}_{C20}$, the $\lambda_2$-eigenfunction of $C_{20}$. It is evident that from the tightness of $\mathbf{u}_{C20}$ follows the tightness of $\mathbf{u}_P$.

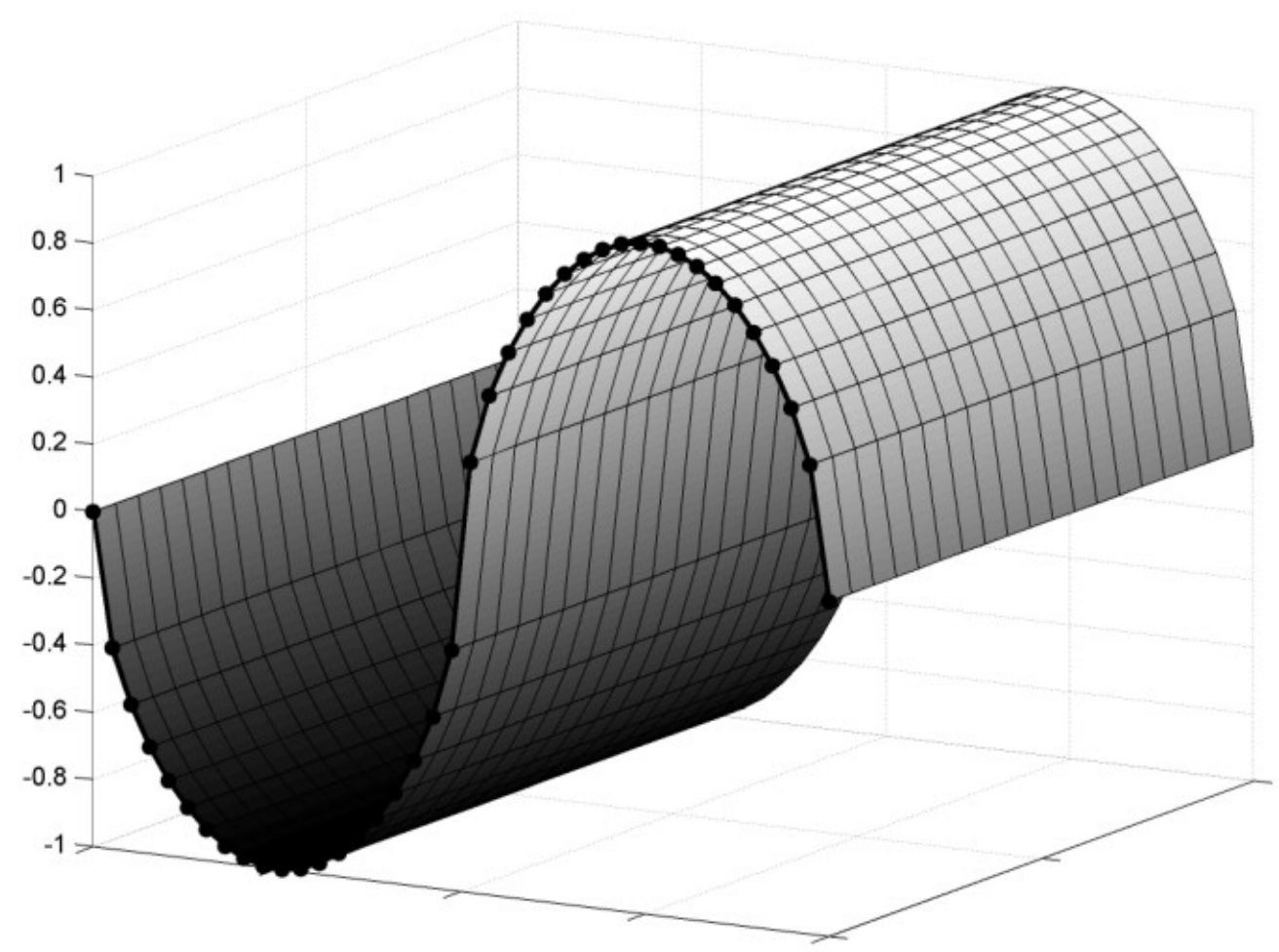

Fig. 3. The $z$ coordinate represents the $\lambda_2$-eigenfunction $\mathbf{u}_P = \mathbf{1} \otimes \mathbf{u}_{C20}$ on the product $C_{10} \square C_{20}$. The 'projection' of the eigenfunction on $C_{20}$ (thick solid line) is $\mathbf{u}_{C20}$, the $\lambda_2$-eigenfunction of $C_{20}$.

Since the Cartesian product is associative, Lemma 3 applies also for a product of any finite number of graphs $P = G_1 \square G_2 \ldots \square G_m$. The $\lambda_2$-eigenspace of $P$ involves only the $\lambda_2$-eigenspaces of the graphs with the least second eigenvalue, $\min\{\lambda_2(G_1), \lambda_2(G_2) \ldots \lambda_2(G_m)\}$, while the rest of the graphs contribute only a factor of $\mathbf{1}$. It is therefore clear from the latter proof that only those $\lambda_2$-eigenspaces need to be tight to assure the tightness of the product's $\lambda_2$-eigenspace. Another corollary of Lemma 3 is that the $\lambda_2$-eigenspaces of a Cartesian product of paths, cycles and complete graphs are tight. *Hamming graphs*, for example, are products of complete graphs and their $\lambda_2$-eigenfunctions are therefore tight.

We note that Lemmas 2 and 3 and their corollaries are only examples for tight $\lambda_2$-eigenspaces. The link between tightness and the absence or presence of saddle points hints for more general tightness criteria. For example, it appears that heterogeneity of the Laplacian weights $\Delta_{ij}$ may induce critical points in the $\lambda_2$-eigenfunctions and thus break tightness.

## 3. Tight polyhedral mappings and the multiplicity of the second eigenvalue.

Next, we show that if the multiplicity of $\lambda_2$ is $m$ then the $m$-dimensional $\lambda_2$-eigenspace spanned by the $m$ eigenfunctions $\mathbf{u}_1$, $\mathbf{u}_2$ …$\mathbf{u}_m$ defines a tight mapping of $G$ into the Euclidean space $\boldsymbol{R}^m$. This therefore sets Banchoff's upper bound on $m$. Figure 4 shows the tight mapping of a triangulated torus of multiplicity $m = 6$.

**Theorem 3**. *If the $\lambda_2$-eigenspace of a Laplacian $\Delta$ on a graph $G$ is tight then its maximal multiplicity $\bar{m}(G)$ is bounded, $\bar{m}(G) \le chr(\gamma(G)) - 1$,* where *$\gamma(G)$ is the genus of the graph and the chromatic number $chr(\gamma)$ is given by Heawood's formula,*

$$chr(\gamma) = \left\lfloor \frac{1}{2}\left(7 + \sqrt{49 - 24\chi}\right) \right\rfloor.$$

**Proof**. To construct the tight mapping we first embed the graph $G$ into a surface $S$ of genus $\gamma(G)$. The embedding is determined by enumerating the faces of the surface. One method to specify the embedding, termed the rotation system, is to list the order of the edges around each of the vertices of $G$ (for details see [27]). The $\lambda_2$-eigenfunctions $\mathbf{u}_1$, $\mathbf{u}_2$ …$\mathbf{u}_m \in \boldsymbol{R}^V$ define a mapping of $S$ into the $m$-dimensional Euclidean space, $\varphi$: $S \to \boldsymbol{R}^m$, by mapping each vertex $v$ to $\varphi(v) = (\mathbf{u}_1(v), \mathbf{u}_2(v) \ldots \mathbf{u}_m(v))$.

Case I: *The mapping preserves edges*, that is any two adjacent vertices $\alpha$, $\beta$ are mapped to different points in $\boldsymbol{R}^m$, $\varphi(\alpha) \neq \varphi(\beta)$. Each edge $(\alpha, \beta)$ is then mapped to the vector stretched between the images of its two endpoint vertices, $\varphi(\alpha)$ and $\varphi(\beta)$, and the faces are mapped to planar polygons determined by their vertices. If the vertices of a certain face are not coplanar then the face is triangulated. Since $\varphi$ preserves edges it is an immersion of $S$ into $\boldsymbol{R}^m$. Every direction in $\boldsymbol{R}^m$ corresponds to a linear combination of the $\lambda_2$-eigenfunctions, which is also tight (definition 1). Thus, by the tightness of the $\lambda_2$-eigenspace the resulting immersion has the two-piece property in any given direction and is therefore tight. By the linear independence of the $\mathbf{u}_i$, the image $\varphi(S)$ is not contained in any hyperplane and the mapping is therefore substantial. Banchoff proved that the maximal substantial dimension $m$ of the mapping is bounded $m \leq chr(\gamma(G)) - 1$ [26, 28, 31], where the chromatic number $chr(\gamma(G))$ is given by Heawood's formula [21]. Since $m$ is also the dimension of the $\lambda_2$-eigenspace it proves the theorem for case I.

Case II: *The mapping does not preserve edges,* that is there exists at least one edge $(\alpha, \beta)$ which is mapped to a point in $\boldsymbol{R}^m$, $\varphi(\alpha) = \varphi(\beta)$. Thus $\varphi$ fails to be an immersion of $G$ and $S$ into $\boldsymbol{R}^m$. To remedy this kind of pathology and make $\varphi$ an immersion (of a modified graph and surface) we apply the following contraction procedure: We identify the vertices $\alpha$ and $\beta$ by contracting the edge $(\alpha, \beta)$. The genus of the resulting contracted graph $G^c$ and the related contracted surface $S^c$ either remains unchanged or decreases. We follow the contraction procedure until all the pathologies are removed. The outcome is a polyhedral immersion, of $G^c$ and $S^c$ into $\boldsymbol{R}^m$.

The "contracted" functions $\mathbf{u}_1^c, \mathbf{u}_2^c \ldots \mathbf{u}_m^c$ that correspond to the immersion are obtained from the original $\lambda_2$-eigenfunctions $\mathbf{u}_i$ by identifying their $\alpha$ and $\beta$ entries for every edge $(\alpha, \beta)$ that is mapped to a point. The linear independence of the $\mathbf{u}_i^c$ follows from that of the $\mathbf{u}_i$, since we only removed entries that were identical in all the $\mathbf{u}_i$. The contracted functions are vectors of length equal to the number of vertices $V$ minus the number of pathological edges. This length is at least $m$ by the linear independence of $\mathbf{u}_i^c$.

To show that $\varphi$ is indeed a tight immersion of $G^c$ and $S^c$, we need to show that the $\mathbf{u}_i^c$ define a tight function-space. The contracted functions $\mathbf{u}_i^c$ span an $m$-dimensional vector-space. Along each direction $i$ in $\boldsymbol{R}^m$ the image $\mathbf{u}_i^c(G^c)$ is equal (as a set) to the image $\mathbf{u}_i(G)$. The tightness of $\mathbf{u}_i^c$ therefore follows from the tightness of $\mathbf{u}_i$. The same is true for any general direction (and linear combination of $\mathbf{u}_i^c$) and the vector-space spanned by $\mathbf{u}_i^c$ is therefore tight. By the linear independence of the $\mathbf{u}_i^c$, the image $\varphi(S^c)$ is not contained in any hyperplane and the mapping is therefore substantial. The genus of the contracted graph and surface is equal or smaller than that of the original graph $\gamma(G^c) \leq \gamma(G)$. By Banchoff's theorem $m \leq chr(\gamma(G^c)) - 1 \leq chr(\gamma(G)) - 1$, where the last inequality follows from the fact that the chromatic number in Heawood's formula is a non-decreasing function of the genus $\gamma$. Since $m$ is the dimension of the $\lambda_2$-eigenspace it proves the theorem for case II. □

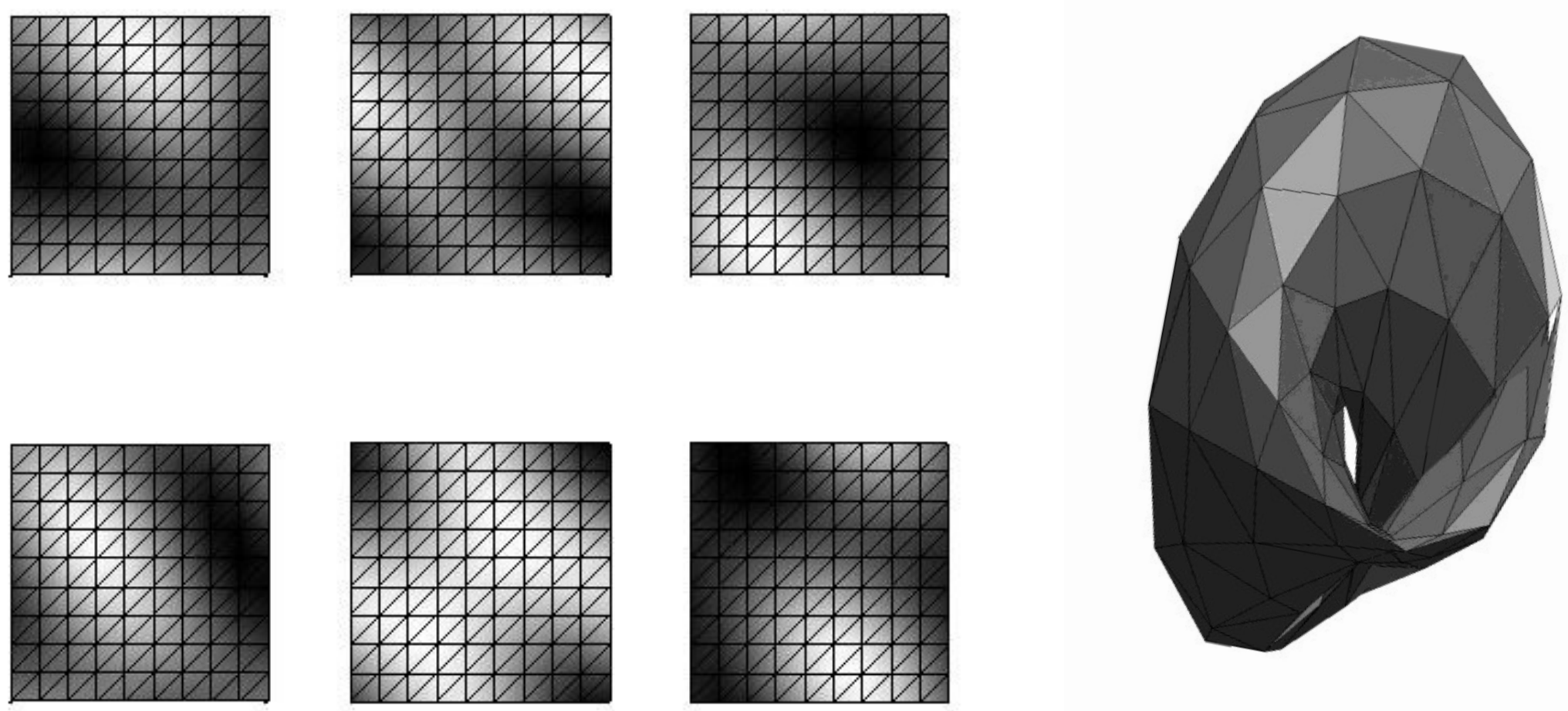

Fig. 4. The six $\lambda_2$-eigenfunctions $\mathbf{u}_1, \mathbf{u}_2 \ldots \mathbf{u}_6$ on a triangulated torus are represented by grayscale (left). The non-diagonal entries of the Laplacian are $\Delta_{ij} = \Delta_{ji} = -1$ if $i$ and $j$ are adjacent, otherwise $\Delta_{ij} = 0$, and the diagonal entries are all $\Delta_{ii} = 6$. The number of eigenfunctions in this case is equal to the conjectured maximal multiplicity $chr(1) - 1 = 6$. The triangulated surface (right) is a three-dimensional projection $\varphi_p(v) = (\mathbf{u}_1(v), \mathbf{u}_2(v), \mathbf{u}_3(v))$ of the six-dimensional tight mapping $\varphi: S \rightarrow \boldsymbol{R}^6$ on three of the $\lambda_2$-eigenfunctions.

Theorem 3 sets Banchoff's or Colin de Verdière's bound on any tight eigenspace, not necessarily that of $\lambda_2$. We also note that Theorem 3 is valid also for Schrödinger operators $H = \Delta + V$, where $V$ is the diagonal potential matrix. Schrödinger operators are the discrete analogues of the general second-order self-adjoint operators. This follows from the applicability of Theorem 1 to Schrödinger operators.

Similar to the definition of $\overline{m}(S)$, one may also define an upper bound $\tilde{r}$ of the $\lambda_2$-multiplicity $m$ over all the graphs of a certain genus $\gamma(G)$ whose $\lambda_2$-eigenspace is tight. By Theorem 3 $\tilde{r}$ $(\gamma)-1$. For any $\gamma$ one can construct the Laplacian over the complete graph whose number of vertices is the chromatic number $K_{chr(\gamma)}$. The genus of this graph is $\gamma(K_{chr(\gamma)}) = \gamma$ and its $\lambda_2$-eigenspace is tight (definition 1). The maximal $m$ for this graph is $\overline{m}(K_{chr(\gamma)}) = chr(\gamma) - 1$, which is achieved when the weights of the Laplacian for all the edges are equal. From this it follows that $\tilde{r}$ $(\gamma)-1$, that is Colin de Verdière's [20] conjecture applies for this class of graphs.